\documentclass[a4paper,11pt]{article}
\pdfoutput=1

\usepackage{jheppub}
\usepackage[export]{adjustbox}
\usepackage{booktabs}
\usepackage[shortlabels]{enumitem}
\usepackage{dsfont}
\usepackage{nicefrac}
\toccontinuoustrue
\def\Tr{\,\mathrm{Tr}\:}
\def\Eq#1{Eq.~(\ref{#1})}
\def\OO{\mathcal{O}}

\begin{document}

\title{Spacelike thermal correlators are almost time-independent}

\author[1]{Simon Caron-Huot,}
\emailAdd{schuot@physics.mcgill.ca}
\affiliation[1]{McGill University, 3600 rue University, Montr\'eal QC
  H3A 2T8 Canada}
\author[2]{Guy D.\ Moore}
\emailAdd{guy.moore@physik.tu-darmstadt.de}
\affiliation[2]{Institut f\"ur Kernphysik,
  Technische Universit\"at Darmstadt, Schlossgartenstra{\ss}e 2,
  64289 Darmstadt, Germany}

\begin{abstract}
{We show that, in relativistic field theories, the thermal correlation
function of $N$ bosonic operators
$\langle \OO_1(x_1,t_1) \OO_2(x_2,t_2)\ldots \OO_N(x_N,t_N)\rangle_T$
at sufficiently spacelike-separated points shows exponentially
weak dependence on the time variables $t_1,\ldots,t_N$, when the
space separations are held fixed.  For classical thermal
field theory, the time dependence vanishes when all points are spacelike separated.}
\end{abstract}

\maketitle
\flushbottom

\section{Introduction}
\label{intro}

In this brief note, we present a property of relativistic
quantum field theories, in equilibrium at nonvanishing temperature
$T$.  Even though we are not aware of any new,
physically relevant applications of this property, we
find it sufficiently striking, and not well known nor
appreciated by the community,
that we thought it appropriate to point it out.

To motivate the discussion, consider the simplest relativistic field
theory: free massless one-component scalar field theory.  The
ordering-averaged two-point correlation function of the scalar field,
in vacuum and expressed in coordinate space, is
\begin{equation}
  \frac{1}{2}  \Big\langle \phi(x,t) \phi(0,0)
  + \phi(0,0) \phi(x,t) \Big\rangle_{T=0} = \frac{1}{4\pi^2 (x^2-t^2)} \,.
\end{equation}
As is well known, the correlation function only depends on spatial
separation $x$ and time difference $t$ through the invariant length
squared $x^2-t^2$, so the correlation function varies as one changes
$t$ at fixed $x$, but not as one moves along a hyperboloid.

\begin{figure}[bh]
  \centering
  \includegraphics[width=0.6\linewidth]{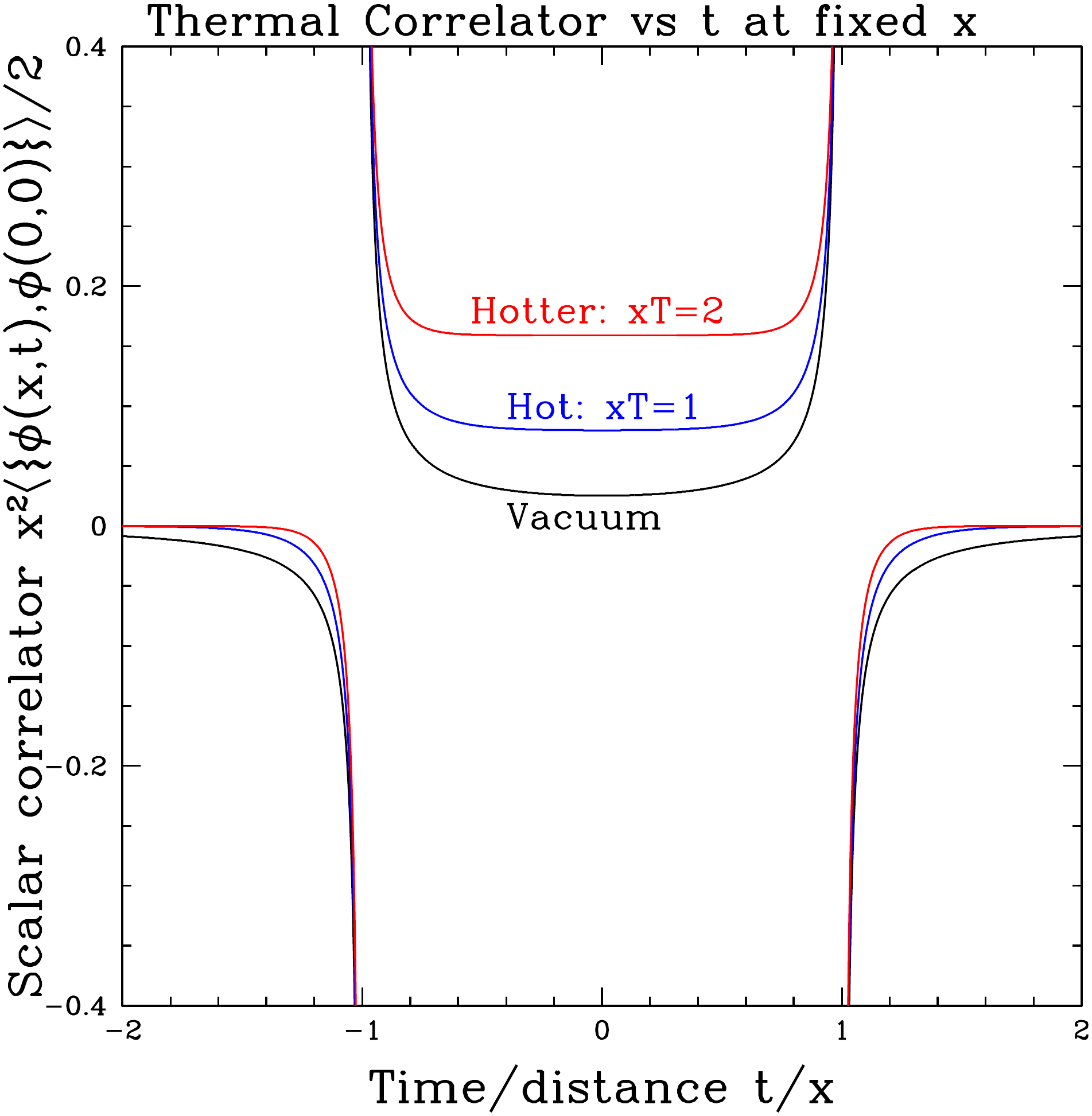}
  \caption{\label{fig:freescalar}
    The free scalar theory correlation function, at fixed separation
    and as a function of time, in vacuum and at two temperatures.  At
    the higher temperature the correlator rapidly approaches zero in
    the timelike region and becomes exponentially flat in the
    spacelike region.}
\end{figure}

The introduction of temperature selects a preferred axis, destroying
this property.  A straightforward calculation (see below) shows that
the correlator becomes
\begin{equation}
  \frac{1}{2}  \Big\langle \phi(x,t) \phi(0,0)
  + \phi(0,0) \phi(x,t) \Big\rangle_{T\neq 0}
  = \frac{T}{8\pi x} \left(
  \frac{1+e^{-2\pi T(x-t)}}{1-e^{-2\pi T(x-t)}}
  +\frac{1+e^{-2\pi T(x+t)}}{1-e^{-2\pi T(x+t)}} \right)  \,.
  \label{eq:freescalar}
\end{equation}
This correlator is illustrated in Figure \ref{fig:freescalar}.
Several features are clear.  Compared to the vacuum, the thermal
correlation function is larger in the
spacelike region and smaller in the timelike region.  It now decays
exponentially at large times.  This property is not generic; it
depends on our choice to examine a field operator in a
free and massless theory.  But another feature turns out to be generic:  the
correlator becomes very flat near $t=0$.  In
fact, closer examination of \Eq{eq:freescalar} shows that the
correlator is \textsl{exponentially} weakly $t$ dependent when both
$x-t$ and $x+t$ are large compared to $1/2\pi T$.
The remainder of the paper
proves that this property holds for arbitrary $N$-point functions in a
general interacting theory, provided that points are sufficiently
spacelike separated from each other.

\section{Points on a hyperplane and perturbation theory}
\label{sec:hyperplane}

In Ref.~\cite{CaronHuot:2008ni} one of us considered the correlation
function of $N$ operators which all lie on a single spacelike
hyperplane, $t = v z$ (or $x^0 = v x^3$) with $-1 < v < 1$.  Although
that reference does not remark on it specifically, the analysis there
shows that the time-dependence of such correlators is exponentially
weak for sufficiently strong spacelike separation, to all orders in
perturbation theory.  Here we will reproduce
this result, and the formalism of \cite{CaronHuot:2008ni}, in a slightly
simpler way.

First consider the free scalar field correlator we already considered,
at finite temperature and at equal time.  It can be computed within
the Matsubara formalism.  We will assume that the reader has some
familiarity with this approach; for a review, see
\cite{Kapusta:2006pm,Bellac:2011kqa,Laine:2016hma}.  The density matrix
$\rho = \frac{1}{Z} e^{-\beta H}$ is expressed as a path integral in
terms of a fictitious Euclidean time $\tau$ which is periodic with
period $\beta = 1/T$; $Z=\Tr e^{-\beta H}$ is the partition function,
as usual.  The equal-time correlation functions of quantum
field theory coincide with the equal-$\tau$ correlation functions in this Euclidean path
integral.  Furthermore, we can consider unequal $\tau$ values if we
want.  Perturbatively this is done by Fourier transforming the $\tau$
direction into a discrete frequency sum -- the Matsubara sum -- with
discrete frequencies $\omega_n = 2\pi n T$.  These behave as heavy modes
with exponentially decaying spatial correlations.

Let us illustrate this for the scalar two-point function we considered above, at unequal $z$ and
$\tau$:
\begin{equation}
  \label{twopointtau}
  \langle \phi(z,\tau) \phi(0,0) \rangle
  = \frac{T}{4\pi z} + T \sum_{n \neq 0} \frac{1}{4\pi z}
  e^{-2\pi |n| T z} \times e^{i 2\pi n T \tau} \,,
\end{equation}
where we have separately written the zero and nonzero frequency
contributions, and $1/4\pi z$ and $e^{-2\pi nTz}/4\pi z$ are the
spatial correlation functions, in 3 dimensions, of a massless field
and a field of mass $m = 2\pi n T$, respectively.
The unequal-time correlation function is just the continuation of the
unequal-$\tau$ correlation function from imaginary to real time,
$\tau \to i t$:
\begin{equation}
  \label{twopointt}
  \langle \phi(z,t) \phi(0,0) \rangle
  = \frac{T}{4\pi z} \left( 1 +
  \sum_{n > 0}  e^{-2\pi n T z} \: e^{-2\pi n T t}
  + \sum_{n > 0} e^{-2\pi nT z} \: e^{2\pi nT t} \right) \,.
\end{equation}
The second sum is the negative $n$ values in \Eq{twopointtau},
relabeled to make $n$ positive.  For $|t|<z$ the sums converge, for
larger $t$ we must rely on analytical continuation.
This leads directly to \Eq{eq:freescalar}.

This generalizes to an interacting theory with nontrivial two-point function $G(z,\tau)$ as follows.
Write the Euclidean frequency-momentum domain correlator
as $G(p,\omega_n)$; then the unequal time correlation function is found by
Fourier transforming to $(z,\tau)$ but replacing $\tau \to i t$:
\begin{equation}
  \label{twopointp}
  \langle \phi(z,t) \phi(0,0) \rangle
  = T \sum_{n} \int \frac{d^3 p}{(2\pi)^3}
  e^{ip_z z}e^{-\omega_n t} G(p,\omega_n) \,.
\end{equation}
The contribution from the $n$'th mode decays like $e^{-m_n |z|}$, where $m_n$, sometimes called the $n$'th screening mass,
parametrizes the position of the first complex-$p$ singularity of $G(p,\omega_n)$. Significantly, $m_n\geq 2\pi |n| T$ for all $n$ in any relativistic theory.\footnote{
The proof is a combination of two standard facts: the Euclidean correlator with positive $n$ coincides with the retarded correlator:
$G(p,\omega_n)=G_R(p,i\omega_n)$ \cite{Bellac:2011kqa};
by causality, the retarded correlator is analytic when ${\rm Im}(p^\mu)$ is future timelike \cite{Haag:1992hx}.
Thus $G(p,\omega_n)$ is analytic when $|{\rm Im}(p)|< |2\pi n T|$ \cite{CaronHuot:2008ni}.}
Therefore, all time-dependent contributions decay at least as fast as $e^{-2\pi T(z-|t|)}$, and the time-independent term dominates away from the light cone (if $m_0<m_n$ for all $n\neq 0$, which we generally expect).
For fermions, only half-integer $n$ contributes so all modes are time-dependent.

Similarly, for an $N$-point function and its Fourier transform in
terms of the first $N-1$ coordinates, we have
\begin{align}
  \label{npointp}
  \langle \OO_1(x_1,t_1) \OO_2(x_2,t_2) \ldots \OO_N(0,0) \rangle
  = & \: T^{N-1} \sum_{n_1,n_2,\ldots n_{N-1}}
  \int \frac{d^3 p_1 \ldots d^3 p_{N-1}}{(2\pi)^{3(N-1)}} \times
   \\ & \:
    G^{\OO_1\OO_2 \ldots \OO_N}(p_1,\omega_{n_1}; p_2,\omega_{n_2}; \ldots)
    e^{i \sum_i \vec{p}_i \cdot \vec{x}_i} e^{-\sum_i \omega_{n_i} t_i} . \nonumber
\end{align}
Here $\omega_{n_i} = 2\pi n_i T$ is the Matsubara frequency associated with the $i$'th operator.
Analyzing this requires more care since the analyticity properties of $G^{\OO_1\OO_2 \ldots \OO_N}$ are less well understood,
and the sum does not converge in all of Minkowski space (it certainly diverges whenever two points become timelike).
Let us thus restrict to the case where all operators lie on a spacelike hypersurface with $t_i = v z_i$.
We can rewrite this expression as
\begin{equation}
  \label{rewrite}
  e^{i \sum_i \vec p_i \cdot \vec z_i} e^{-\sum_i \omega_{n_i} t_i}
  \to e^{i \sum_i  \tilde p_i \cdot \vec z_i} \,, \quad
  \tilde p_z \equiv p_z + i v \omega_n \,.
\end{equation}
This imaginary shift in $p_z$ can be removed by shifting the
integration variable by $p_z \to p_z' \equiv p_z - i v \omega_n$.  This
shift never encounters singularities in $G^{\OO_1 \ldots \OO_N}$,
provided $|v|<1$, as discussed in \cite{CaronHuot:2008ni}.  After this
shift, all reference to time is hidden in this shift of the momentum
integration variable, or equivalently, in a shift to the $p_z$
arguments of the momentum-space correlation function:
$G^{\OO_1 \OO_2 \ldots \OO_N}(p_1',\omega_{n_1}; p_2',\omega_{n_2}; \ldots)$.
That is, we can formulate the Matsubara formalism for the case of a
spacelike hyperplane by replacing $p_z \to p_z'=p_z - iv\omega_n$ on all
propagators (and, in a derivative-coupled theory, all
momentum-dependent vertices).  This is equivalent to the formulation
of Ref.~\cite{CaronHuot:2008ni}, which arrived at this result somewhat
differently.

\begin{figure}[bht]
  \centering
  \includegraphics[width=0.5\linewidth]{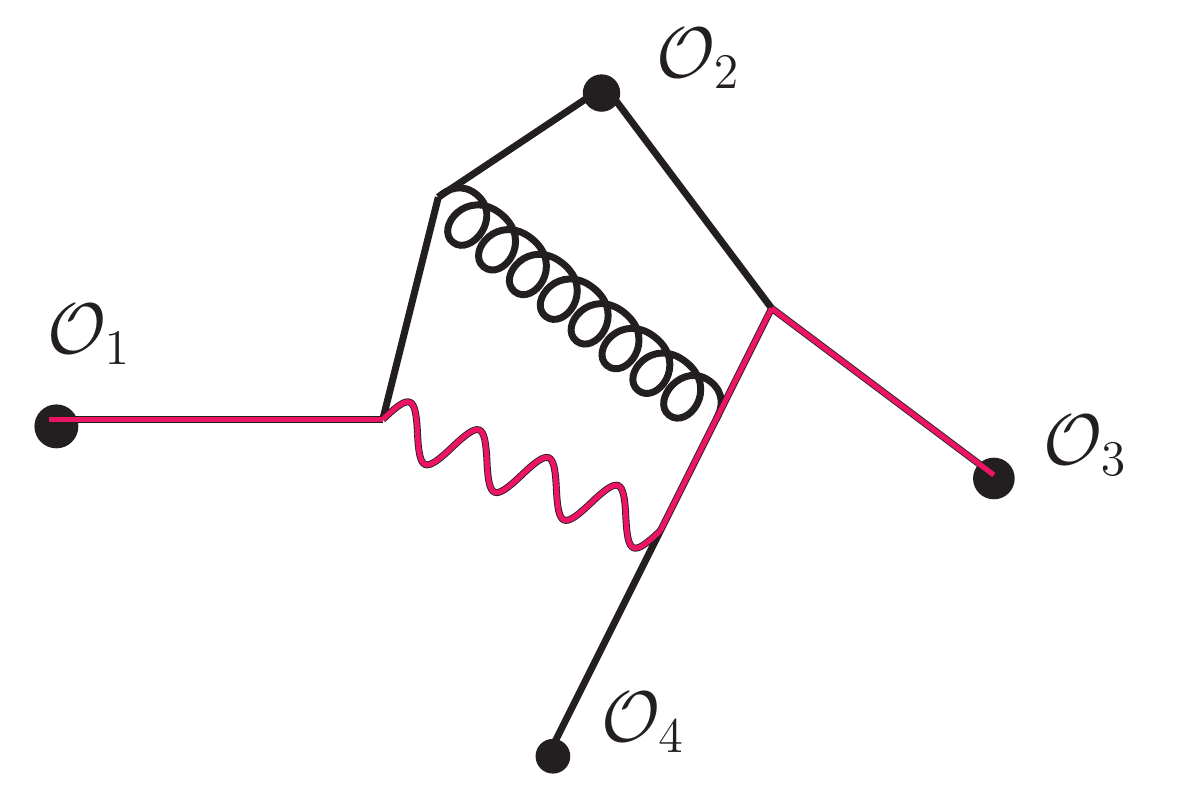}
  \caption{\label{genericdiagram}
  A typical diagram contributing to some 4-operator correlation
  function.  The red lines indicate a possible path for a nonzero
  Matsubara frequency to cross the diagram.}
\end{figure}

It is instructive to consider a generic diagram contributing to the correlation function in pertubation theory, as
depicted in Figure \ref{genericdiagram}.  If all the external lines
have zero Matsubara frequency (all $n_i = 0$ in \Eq{npointp}) then
there is no shift in the $p_z$ variables.  Such a shift can still
occur in a loop momentum, but no singularities are encountered if we
shift the loop integration variable back to real $q_z$, restoring
precisely the equal-time perturbative expansion.  Therefore, the
contribution from zero external Matsubara frequencies, for
$|v|<1$, is identical to that at $v=0$, that is, equal time.
Any time dependence must arise when one or more external Matsubara
frequencies are nonzero.

But if an external Matsubara frequency is nonzero, then momentum
conservation at all vertices requires that this frequency flows
through the diagram to another external line (to another operator).
There are a series of propagators carrying this nonzero frequency from
the incoming to the outgoing operator, as indicated in Figure
\ref{genericdiagram} by the series of red lines.  Each
nonzero-frequency propagator connects spacelike separated points, and
is suppressed by $\exp(-2\pi |n|T(|\Delta x| \mp v \Delta z))$.  On a
spacelike hyperplane, the
sum of the spacelike lengths of a series of segments is at least as
large as the spacelike length of the straight-line path; so this leads to a
suppression of at least $\exp(-2\pi |n|T(|x|\mp v|t|))$.  Therefore the time
dependence is exponentially suppressed provided that all pairs of
external points have $|x|-|t| \gg 1/2\pi T$.
In this argument it was essential that an Euclidean formalism applies when
all operators lie on a spacelike hypersurface.

\section{General case}
\label{sec:general}

So far we have shown that $N$-point functions of bosonic operators in
relativistic, thermal quantum field theory have exponentially small
time dependence to all orders in perturbation theory, provided that
\begin{itemize}
\item
  all coordinates are ``strongly'' spacelike separated,
  $|\Delta x|-|\Delta t| \gg 1/2\pi T$, and
\item
  all coordinates lie on a spacelike hyperplane.
\end{itemize}
But does this hold fully nonperturbatively?  And is there anything
special about spacelike hyperplanes, or does this hold more generally,
provided that some operator is spacelike separated from others?

To address this, consider $N$ operators $\OO_1(x_1,t_1)$ through
$\OO_N(x_N,t_N)$.
We will assume that $x_1$ is far enough from other points that there is some range of $t_1$
values where $(x_1,t_1)$ is spacelike separated from every other $(x_i,t_i)$.
Now consider the dependence on $t_1$ of the following two functions:
\begin{align}
  \label{def:GSGA}
  G_S(t_1) \equiv & Z^{-1} \frac{1}{2} \Tr e^{-\beta H} \left(
  \OO_1(x_1,t_1) \OO_{\mathrm{others}}
  +  \OO_{\mathrm{others}} \OO_1(x_1,t_1) \right)
  \,, \nonumber \\
  G_\sigma(t_1) \equiv & -i Z^{-1} \Tr e^{-\beta H} \left(
  \OO_1(x_1,t_1) \OO_{\mathrm{others}}
  -  \OO_{\mathrm{others}} \OO_1(x_1,t_1) \right) \,.
\end{align}
Here $\OO_{\mathrm{others}}$ is the product of the other operators, taken in some arbitrary but fixed order.
We recognize
$G_S(t_1)$ as the symmetrized function, and $G_\sigma(t_1)$ as the
commutator or spectral function.  We can Fourier transform each
function, and in Fourier space they obey a KMS relation, derived in a standard way by commuting $\OO_1$ across $e^{-\beta H}$:%
\footnote{%
  Here is a quick derivation of the KMS relation.  Write
  $\OO_1(t) = e^{iHt} \OO_1 e^{-iHt}$ and insert a complete set of
  (energy eigen)states $\sum_m |m\rangle \langle m|$ to its left and
  $\sum_n |n \rangle \langle n|$ to its right.  Then the lefthand side
  of \Eq{KMS} becomes $\sum_{mn} \int dt e^{i\omega t}
  e^{i(E_m-E_n)t} e^{-\beta E_m} \langle m|\OO_1|n\rangle
  \langle n|\OO_{\mathrm{others}}|m \rangle$.  The right-hand side is
  the same except with $e^{-\beta E_m} \to e^{-\beta E_n}$.
  Performing the time integral introduces $\delta(\omega+E_m-E_n)$,
  allowing us to rewrite $e^{-\beta E_m} = e^{\beta \omega} e^{-\beta E_n}$.
  The relation then follows.}
\begin{equation}
  \label{KMS}
  G^>(\omega) \equiv \int dt\: e^{i\omega t}
  \Tr e^{-\beta H} \OO_1(t) \OO_{\mathrm{others}}
= e^{\beta \omega}
  \int dt\: e^{i\omega t} \Tr e^{-\beta H} \OO_{\mathrm{others}}
    \OO_1(t)
    \equiv e^{\beta \omega} G^<(\omega)
\end{equation}
and therefore
\begin{align}
  \label{GSGArelation}
  G_S(\omega) &= \frac{e^{\beta \omega}+1}{2} G^<(\omega) \,, \quad
  G_\sigma(\omega) = -i( e^{\beta \omega} -1 ) G^<(\omega) \,, \quad
  \mbox{and hence}
  \nonumber \\
  G_S(\omega) &= -i \left( n_b(\omega)+\frac 12 \right) G_\sigma(\omega),
\end{align}
where $n_b(\omega) = 1/(e^{\omega/T} - 1)$ is the Bose-Einstein
distribution function.
These statements are nonperturbative, relying only on energy
conservation and the form of the thermal density matrix.

In the high temperature or low-frequency limit we can approximate this
as
\begin{equation}
  \label{HighT}
  G_S(\omega) \simeq -i \frac{T}{\omega} G_\sigma(\omega) \qquad
  \mbox{or, in the time domain,} \qquad
  G_\sigma(t) = -\frac{1}{T} \, \frac{d}{dt} G_S(t) \,.
\end{equation}
In this approximation, the spectral function is $(-1/T)$ times the
time derivative of the symmetrized function.  But causality ensures
that the spectral function vanishes unless $(x_1,t_1)$ is null- or
timelike- related to at least one of the other operators.  Therefore,
in the time range where $(x_1,t_1)$ is spacelike separated from all others, the
correlator $G_S(t)$ is independent of the time $t_1$.
This strict high-temperature limit is equivalent to the classical
field approximation.

Relaxing the high-temperature approximation, we can Fourier transform
\Eq{GSGArelation} to find a kind of dispersive representation:
\begin{equation} \label{GS disp}
  G_S(t_1) = T \int dt_1' G_\sigma(t_1') \left(
  \frac{1+e^{-2\pi T(t_1'-t_1)}}{1-e^{-2\pi T(t_1'-t_1)}} \right) \,.
\end{equation}
The quantity in parenthesis, which is the Fourier transform
of $-i\left(n_b(\omega)+\frac{1}{2}\right)$, is constant up to
exponentially small corrections away from $(t_1'-t_1)=0$.  Therefore, if
$G_\sigma(t_1')$ vanishes in some window around $t_1'=t_1$, the time dependence of
$G_S(t_1)$ is exponentially small.  But
$G_\sigma(t_1')$ is nonvanishing only if $(x_1,t_1')$ is null-
or timelike- separated from at least one other operator.  Therefore,
when $(x_1,t_1)$ is strongly spacelike separated from all other points, the time dependence of $G_S(t_1)$ is exponentially suppressed by $2\pi T$ times the distance to the closest lightcone.
Note that a stronger suppression is possible, as illustrated above in the two-point case where the exponent features $m_1\geq 2\pi T$.

\section{Discussion}
\label{sec:discussion}

We have shown that spacelike correlation functions in relativistic,
thermal field theories have exponentially weak dependence on the
relative times of the operators, suppressed by at least $\exp(-2\pi T|x-t|)$,
with $|x-t|$ the time separation from the nearest light cone and $T$ the temperature.  This
property does not depend on perturbation theory; indeed, it follows
only from causality and the KMS relation.

In a strong coupling scenario, this result may not be too surprising.
Then one expects static correlations to decay with a mass $m_0\sim 2\pi T\times O(1)$ \cite{Bak:2007fk},
so they may already be small in regimes where they dominate over time-dependent correlations.
Similarly, our result has no implications for equilibrium hydrodynamics (thermal hydrodynamical fluctuations); while hydrodynamical correlators are
highly nontrivial in the timelike region, they generally vanish at
spacelike separation.
In theories near a critical point, there are
strong long-range fluctuations, and in this case our result provides
nontrivial information about the time dependence of correlation
functions.  Namely, at a separation $x \lesssim \xi$ with $\xi$ the
correlation length, there is a time scale $t \in [-x,x]$ over which
correlators do not evolve.  This is unexpected.  We emphasize,
however, that it does not contradict any standard results regarding
dynamical critical fluctuations, which pertain to time scales of
order $t \sim x^z T^{z-1}$ with $z>1$ the dynamical critical exponent
\cite{Hohenberg:1977ym}
and $1/T$ an estimate for the UV scale below which criticality
arguments become applicable.  Therefore, the standard analysis of
dynamical critical scaling involves time scales which are
parametrically longer than the time scale where our result is valid.

We find the generality of our results surprising, and thought that
they were worth presenting, even though we are not yet aware of
applications, other than those to jet quenching at weak coupling which
are explored in
Ref.~\cite{CaronHuot:2008ni}.\\

\noindent {\bf Acknowledgements}\\
Work of SCH is supported by the Simons Collaboration on the
Nonperturbative Bootstrap, the Canada Research Chair program and the
Sloan Foundation.
GDM acknowledges the support by the Deutsche Forschungsgemeinschaft
(DFG, German Research Foundation) through the CRC-TR 211
``Strong-Interaction Matter under Extreme Conditions'' -- project
number 315477589 - TRR 211.

\bibliographystyle{JHEP}
\bibliography{refs}

\end{document}